\documentclass[twocolumn,showpacs,preprintnumbers,amsmath,amssymb]{revtex4}

\usepackage{graphicx}
\usepackage{dcolumn}
\usepackage{bm}

\begin{document}


\title{PREDICTION OF THE IN-GAP STATES ABOVE THE TOP OF THE VALENCE BAND IN THE UNDOPED INSULATING CUPRATES DUE TO SPIN-POLARON EFFECT}

\author{S.G. Ovchinnikov}
 \email{sgo@iph.krasn.ru}
\author{A.A. Borisov}
\author{V.A. Gavrichkov}
\author{M.M. Korshunov}
\affiliation{%
L.V. Kirensky Institute of Physics Siberian Branch of Russian Academy of Science, Krasnoyarsk, 660036, Russia
}%

\date{\today}

\begin{abstract}
In the framework of the generalized tight binding method we have  calculated the quasiparticle band structure and the spectral functions of the undoped cuprates like $La_2 CuO_4$, $Sr_2CuO_2Cl_2$ etc. Due to spin fluctuations the in-gap state appears above the top of the valence band in the undoped antiferromagnetic insulator similar to in-gap states induced by hole doping. In the ARPES experiments the in-gap states can be detected as weak low energy satellites.
\end{abstract}

\maketitle

\section{Introduction}

The key issue to understand the nature of high-temperature superconductivity 
in the cuprates is the evolution of the electronic structure from an 
antiferromagnetic insulator to a superconductor with hole doping. The 
appearance of the in-gap states above the top of the valence band in 
slightly doped cuprates has been found experimentally [1-4]. In the metallic 
underdoped regime the ARPES measurements [5] reveal the concentration 
dependent band structure of Bi-2212. With improving the ARPES resolution 
recently the formation of the new quasiparticle states at the transition 
from insulator to metal in $La_{2 - x} Sr_x CuO_4 $ has been found [6, 7].

The formation of the in-gap states with doping has been obtained 
theoretically in the numerical studies of small cluster in the framework of 
the $t - J$ model, Hubbard model, and the 3-band $p - d$ model [8-11]. The 
band structure calculations of the $CuO_2 $ layer in the framework of the 
multiband $p - d$ model by the generalized tight-binding (GTB) method [12] 
with account for strong electron correlations have revealed the unusual 
in-gap state at the top of the valence band with zero spectral weight for 
undoped insulator that acquire the dispersion and non-zero spectral weight 
with hole doping [13]. In all models of strongly correlated electrons the 
hopping of hole in the antiferromagnetic background is renormalized by spin 
fluctuations. To clarify the origin of the in-gap state we have studied in 
this paper a spin-polaron effect both analytically in the framework of the 
$t - t' - J$ model and numerically by the GTB method similar to [13]. We 
have found that the spin fluctuations provides non-zero spectral weight and 
dispersion of the in-gap state as well as hole doping does -- even without 
doping the spin excitations that are present in the antiferromagnetic state 
due to the quantum spin fluctuations at all temperatures including $T=
0$ results in the non-zero in-gap spectral weight and dispersion above the 
top of the valence band. This state can be detected by the ARPES measurement 
as a weak satellite at the low energy shoulder of the main peak. Moreover 
the concentration of spin fluctuations $n_{sf} $ increases with temperature, 
and we expect the growth of the in-gap spectral weight $\sim n_{sf} $.

\section{GTB method results \label{section:2}}

A dispersion equation of the GTB method for the quasiparticle band structure 
of the $CuO_2 $ layer looks like [13]
\begin{eqnarray}
\Bigl\| &&\left( E - \Omega ^A_m  \right)\delta_{mn} \nonumber \\
&&- 2F_\sigma ^A \left(m \right)\sum\limits_{\lambda {\lambda }'} {\gamma _{\lambda \sigma }^\ast \left( m \right)T_{\lambda \lambda '}^{AB} \left( \vec {k} \right)\gamma 
_{{\lambda }'\sigma } \left( n \right)} \Bigl\| = 0
\label{eq1}
\end{eqnarray}
Here $m$ is a quasiparticle band indexes given by a pair $\left( {p,q} 
\right)$ of the initial and final multielectron configurations $E_p (n + 1)$ 
and $E_q (n)$, $\Omega _m = E_p (n + 1) - E_q (n)$ is a local excitation 
energy. The local excitation $\left| q \right\rangle \to \left| p 
\right\rangle $ is described by the Hubbard operator $X^{pq} = \left| p 
\right\rangle \left\langle q \right|$, a filling factor $F(m) = \left\langle 
{X^{pp}} \right\rangle + \left\langle {X^{qq}} \right\rangle $. Two magnetic 
sublattices are denoted by indexes $A$ and $B$, and $\sigma $ is a spin 
projection. The interatomic hopping is $T_{\lambda \lambda '}^{AB} $, where 
the single hole basis set $\lambda $ includes 5 orbitals: copper $d(x^2 - 
y^2)$, $d(3z^2 - r^2)$, in-plane oxygen $p(x)$, $p(y)$ and apical oxygen 
$p(z)$, $\gamma _{\lambda \sigma } (m)$ is a parameter of a single hole 
annihilation operator in term of the Hubbard operators
\begin{equation}
\label{eq2}
a_{\lambda \sigma } = \sum\limits_m {\gamma_{\lambda \sigma } (m)X^m} 
\end{equation}
The local multielectron energies and parameters $\gamma _{\lambda \sigma } 
(m)$ are obtained after the exact diagonalization of the multiband $p - d$ 
model Hamiltonian for the unit cell. In our case the unit cell is $CuO_2 $ 
cluster for $La_2 CuO_4 $ and $CuO_4 Cl_2 $ for $Sr_2 CuO_2 Cl_2 $. The 
similar equation has been known long ago for the nondegenerate Hubbard model 
as the Hubbard I solution and has been used recently to study magnetic 
properties of transition metals [14, 15].

The essential for undoped cuprates multielectron configurations are 
$d^{10}p^6$ (vacuum state $\left| 0 \right\rangle $ in a hole 
representation), single-hole configurations $d^9p^6$, $d^{10}p^5$, and 
two-hole configurations $d^8p^5$, $d^9p^5$, $d^{10}p^4$, $d^{10}p^5p^5$. The 
minimal energy in the single-hole sector of the Hilbert space has the 
$b_{1g} $ molecular orbital, and in the two-hole sector the $^1A_{1g} $ 
singlet that besides Zhang-Rice singlet contains several more local 
singlets. A staggered magnetic field split $b_{1g} $ levels by spin:
\begin{equation}
\label{eq3}
\varepsilon _{A\sigma } = \varepsilon _1 - \sigma h,
\quad
\varepsilon _{B\sigma } = \varepsilon _1 + \sigma h.
\end{equation}

The top of the valence band is given by the quasiparticles with $m=1$: $X_A^1 = 
\left| {b_{1g, \uparrow } } \right\rangle \left\langle {^1A_{1g} } \right|$ 
and $X_B^1 = \left| {b_{1g, \downarrow } } \right\rangle \left\langle 
{^1A_{1g} } \right|$, as usually there is spin degeneracy of the band in the 
antiferromagnetic state. The occupation number $n_p \equiv \left\langle 
{X^{pp}} \right\rangle $ are calculated self-consistently via the chemical 
potential equation. In the mean-field Hubbard I approximation the solution 
of this equation for the hole-doped cuprates with hole concentration $n_h = 1 
+ x$ is given by
\begin{equation}
\label{eq4}
n_{1 \uparrow } \equiv n_{A \uparrow } (b_{1g} ) = 1 - x, n_{1 \downarrow } = 0, n_2 \equiv n(^1A_{1g} ) = x.
\end{equation}

For the band $m = 1$ we get $F_{A \uparrow } (1) = 1$ while for the band $m 
= 2$ with $X_A^2 = \left| {b_{1g, \downarrow } } \right\rangle \left\langle 
{^1A_{1g} } \right|$ the filling factor is $F_{A \downarrow } (2) = x$. The 
quasiparticle spectral weight is proportional to the filling factor, thus it 
is the band $m=2$ that forms the in-gap state. In the limit $x \to 0$ its 
spectral weight is zero, when $x \ne 0$ this band acquires both dispersion 
and nonzero spectral weight. The corresponding concentration dependent bands 
structure has been obtained for $La_{2 - x} Sr_x CuO_4 $ in [13] and the 
chemical potential $\mu (x)$ dependence, the Fermi surface evolution with doping have 
been studied in [16].

To go beyond the mean-field Hubbard I approximation one has to calculate 
single-loop diagrams for the self-energy [17]. In the ferromagnetic or 
antiferromagnetic state the most important contribution is given by loops 
with spin-wave excitations [18] (a spin-polaron effect). According to [18], 
the main effect of the spin excitations is given by the spin-wave 
renormalization of the multielectron configuration's occupation numbers, so 
instead of (\ref{eq4}) one gets
\begin{equation}
\label{eq5}
n_{1 \uparrow } = (1 - x)(1 - n_{sf} ), n_{1 \downarrow } = (1 - x)n_{sf}, n_2 = x,
\end{equation}

\noindent
where $n_{sf} $ is the occupation of the spin-minority level and it 
determines the spin-fluctuation decrease of the sublattice magnetization
\begin{equation}
\label{eq6}
\left\langle {S_A^z } \right\rangle = (1 - x)(1 / 2 - n_{sf} ).
\end{equation}

Concentration of the spin fluctuations is equal to $2n_{sf} $.

\begin{figure}
\includegraphics[width=\linewidth]{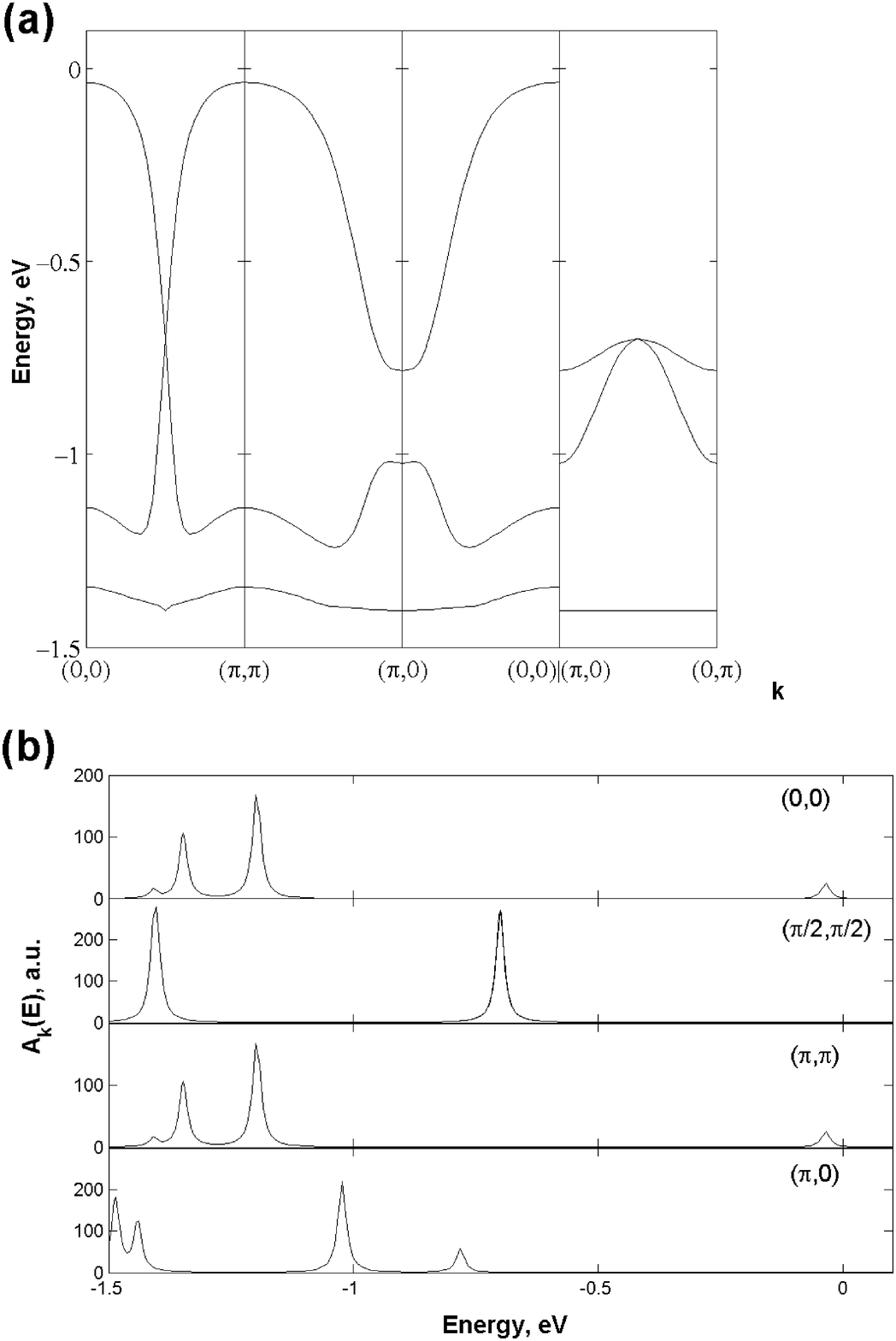}
\caption{\label{fig1} The quasiparticle band structure (a) and the spectral function (b) of
the undoped $La_2 CuO_4 $ with a spin-fluctuations $n_{sf} = 0.2$ calculated by the GTB method. The Fermi level is above all bands shown here}
\end{figure}

Thus the filling factors for the valence band $F(1) = 1 - n_{sf} $, and for 
the in-gap states $F(2) = x + n_{sf} $. It means that the spin-polaron 
effect results in the non-zero spectral weight of the in-gap states even for 
undoped cuprates $La_2 CuO_4 $ and $Sr_2 CuO_2 Cl_2 $. The quasiparticle 
band structure and the spectral function for the undoped $La_2 CuO_4 $ with 
$n_{sf} = 0.1$ are given in the Fig.~\ref{fig1}. Here the lowest band is formed by 
hole hopping via 2-hole triplet $^3B_{1g} $ state -- this aspect was 
discussed in [13, 19]. The next band ($m = 1)$ is the top of the valence 
band without spin fluctuations with a maximum at $k = \left( {\pi / 2,\pi / 
2} \right)$. The upper band ($m=2$) formed by the dispersion of the in-gap states. 
Despite of its width each state has a low spectral weight as seen in the 
Fig.~\ref{fig1}b and the total number of states in this in-gap band (without doping) 
is equal to $n_{sf} $. The appearance of such non-Fermi liquid states is the 
direct effect of strong electron correlations. The maximal spectral weight 
of the in-gap state is near $\left( {\pi ,0} \right)$ point of the Brillouin 
zone (BZ). At the $\left( {\pi / 2,\pi / 2} \right)$ point the two bands are 
degenerate, and we cannot separate the contribution of the in-gap band to 
the spectral function $A_k (E)$.

\section{$t-t'-J$ model treatment}

To clarify the properties of the in-gap band we study the spin-polaron 
effect in the $t - t' - J$ model, which is an effective low 
energy model for the multiband $p - d$ model [20], with the Hamiltonian:
\begin{eqnarray}
H_{t - J} &=& \left( {\varepsilon _1 - \mu } \right)\sum\limits_{f,\sigma } {X_f^{\sigma \sigma } } + \sum\limits_{ < f,g > ,\sigma } {t_{fg} X_f^{\sigma 0} X_g^{0\sigma } } \nonumber \\
&+& \sum\limits_{ < f,g > } {J_{fg} \left( {{\rm {\bf S}}_f {\rm {\bf S}}_g - \frac{1}{4}n_f n_g } \right)},
\label{eq7}
\end{eqnarray}

\noindent
where ${\rm {\bf S}}_f $ are spin operators and $n_f $ are number of 
particle operator, $t_{fg} $ and $J_{fg} $ are the hopping and exchange 
integrals correspondingly. In the Hubbard I 
approximation it is easy to obtain in the undoped case the following 
intrasublattice and intersublattice Green functions ($f \in A,g \in B)$
\begin{eqnarray*}
\left\langle {\left\langle {X_f^{0\sigma } } \mathrel{\left| {\vphantom 
{{X_f^{0\sigma } } {X_{f}'^{\sigma 0} }}} \right. \kern-\nulldelimiterspace} 
{X_{f}'^{\sigma 0} } \right\rangle } \right\rangle _E = 
\frac{2}{N}\sum\limits_k {G_{k\sigma }^{AA} } (E)e^{i\vec {k}(\vec {f} - 
{\vec {f}}')}, \\
\left\langle {\left\langle {X_g^{0\sigma } } \mathrel{\left| {\vphantom 
{{X_g^{0\sigma } } {X_{f}'^{\sigma 0} }}} \right. \kern-\nulldelimiterspace} 
{X_{f}'^{\sigma 0} } \right\rangle } \right\rangle _E = 
\frac{2}{N}\sum\limits_k {G_{k\sigma }^{BA} } (E)e^{i\vec {k}(\vec {g} - 
{\vec {f}}')}.
\end{eqnarray*}
Matrix Grin function in momentum space could be written as:
\begin{eqnarray}
\hat {G}_{k\sigma } &=& \left( {{\begin{array}{*{20}c}
 {G_{k\sigma }^{AA} } \hfill & {G_{k\sigma }^{AB} } \hfill \\
 {G_{k\sigma }^{BA} } \hfill & {G_{k\sigma }^{BB} } \hfill \\
\end{array} }} \right) \nonumber \\
&=& \frac{1}{D}\left( {{\begin{array}{*{20}c}
 {n_{A\sigma } \left( {E - \varepsilon _{k\sigma }^B } \right)} \hfill & 
{n_{A\sigma } n_{B\sigma } t_k^B } \hfill \\
 {n_{A\sigma } n_{B\sigma } t_k^B } \hfill & {n_{B\sigma } \left( {E - 
\varepsilon _{k\sigma }^A } \right)} \hfill \\
\end{array} }} \right),
\label{eq8}
\end{eqnarray}

\noindent
where $D = \left( {E - E_{k\sigma }^ + } \right)\left( {E - E_{k\sigma }^ - } \right)$, $\varepsilon _{k\sigma }^\alpha = \left( {\varepsilon _1 - \mu } \right) - 
\left( {J_0^B - t_k^A } \right)n_{\alpha \sigma } - J_0^A n_{\alpha \bar 
{\sigma }}$, with $\alpha = A,B$.

Here $t_k^B $ and $t_k^A $ ($J_0^B $ and $J_0^A )$ are the hoppings 
(exchanges) in momentum space between different and the same sublattices 
correspondingly. In simple case of next-nearest-neighbors approximation we 
have
\[
t_k^B = 2t\left( {\cos k_x + \cos k_y } \right),
\quad
t_k^A = 4t'\cos k_x \cos k_y ,
\]
\[
J_0^B = 4J,
\quad
J_0^A = 4J',
\]

\noindent
with primed values corresponding to next-nearest hoppings and exchanges. 
The occupation factors of one particle state with different spin projections are denoted by $n_{A\sigma}$ and $n_{B\sigma}$. In the mean field Hubbard I 
approximation $n_{A \uparrow } = (1 - n_{sf} )$ and $n_{A \downarrow } = 0$ at 
$T = 0$. Using the same arguments as in the Section \ref{section:2}, we go beyond 
the Hubbard I approximation by renormalization the occupation numbers with 
spin fluctuations. It results in the undoped $La_2 CuO_4 $ in
\begin{equation}
\label{eq9}
n_{A \uparrow } = (1 - n_{sf} ), n_{A \downarrow } = n_{sf}, n_{B\sigma } = n_{A\bar {\sigma }}.
\end{equation}

The condition $D = 0$ gives two branches of quasiparticle spectrum:
\begin{equation}
E_{k \uparrow }^\pm = \varepsilon _1 - \mu + \frac{1}{2}\left[ {t_k^A - 
J_0^B - J_0^A \pm \sqrt {\beta_k} } \right].
\label{eq10}
\end{equation}
\noindent
where 
\[
\beta_k = \left( {t_k^A - J_0^B + J_0^A } \right)^2\left( {1 
- 2n_{sf} } \right)^2 + 4\left( {t_k^B } \right)^2\left( {1 - n_{sf} } 
\right)n_{sf}.
\]

If we set concentration of the magnons to zero we immediately get one 
dispersionless state and one dispersive state with dispersion governed by 
hoppings between different sublattices:
\begin{equation}
\label{eq11}
\left.{E_{k \uparrow }^ + } \right|_{n_{sf} = 0} = \varepsilon _1 - \mu - J_0^A, 
\left.{E_{k \uparrow }^ - } \right|_{n_{sf} = 0} = \varepsilon _1 - \mu + t_k^A - J_0^B.
\end{equation}

If the values of inter-sublattice hoppings and exchange are small then the 
difference between two energy levels is of order $J$: 
\begin{equation}
\label{eq12}
\left. {\Delta E} \right|_{n_{sf} = 0} = \left. {\left( {E_{k \uparrow }^ + 
- E_{k \uparrow }^ - } \right)} \right|_{n_{sf} = 0} \approx 4J.
\end{equation}

\begin{figure}
\includegraphics[width=\linewidth]{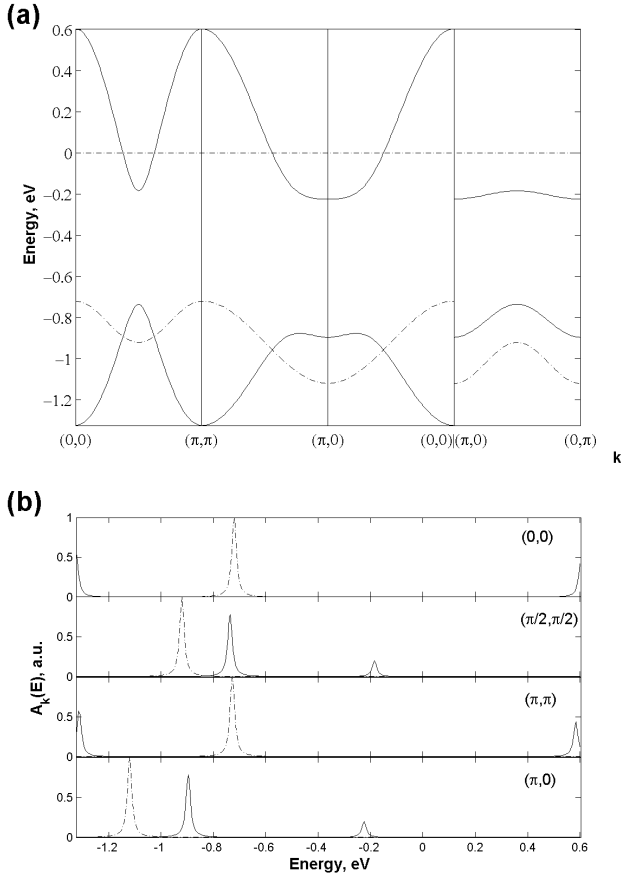}
\caption{\label{fig2} Quasiparticle dispersion (a) and spectral function peaks (b) in the 
$t - J$ model for $x = 0$ and $n_{sf} = 0.2$ (solid lines) and $n_{sf} = 0$ (dot-dashed lines)}
\end{figure}

In Fig.~\ref{fig2} the quasiparticle dispersions corresponding to equations (\ref{eq10}) and 
(\ref{eq11}) are shown. Parameters were taken from effective low-energy model [20] 
of multiband $p-d$ model and equal to $t = - 0.587$, $t' / t = - 0.085$, $J / t = 
0.392$, $J' / t = 0.00037$. The value $n_{sf} $ is calculated in the 
effective quasi-two-dimensional Heisenberg antiferromagnetic model, $n_{sf} 
= 0.2$ for typical in $La_2 CuO_4 $ ratio $10^{ - 5}$ of the interplane and intraplane 
exchange parameters [21].

The distance between two spectrum branches for non-zero concentration of the 
magnons is less then distance (\ref{eq12}) for $n_{sf} = 0$ by factor proportional 
to $\left( {1 - 2n_{sf} } \right) = 2\left\langle {S^z} \right\rangle $.

The lower quasiparticle branch for $n_{sf} = 0.2$ (but not for $n_{sf}=0$!) clearly resembles dispersion obtained in self-consistent Born approximation [22] and GTB method [13]. This proves that two different approaches to treat spin fluctuations lead to the similar results.

Introducing energy difference $\Delta E_k \equiv E_{k \uparrow }^ + - E_{k 
\uparrow }^ - $ we can write down spectral functions $A_{k\sigma } \left( E 
\right) = - \frac{1}{\pi }Im\left[ {Sp\hat {G}_{k\sigma } } \right]$ in the 
form:
\begin{eqnarray}
A_{k \uparrow } \left( E \right) = u_k^2 \delta \left( {E - E_{k \uparrow }^ 
+ } \right) + v_k^2 \delta \left( {E - E_{k \uparrow }^ - } \right), \\
A_{k \downarrow } \left( E \right) = v_k^2 \delta \left( {E - E_{k 
\downarrow }^ + } \right) + u_k^2 \delta \left( {E - E_{k \downarrow }^ - } 
\right),
\label{eq13}
\end{eqnarray}

\noindent
where 
\[
u_k^2 = \frac{1}{2} - \left( {1 - 2n_{sf} } \right)^2\frac{\left( {J_0^B - 
J_0^A - t_k^A } \right)}{2\Delta E_k },
\quad
v_k^2 = 1 - u_k^2 .
\]

Obviously, for $n_{sf} = 0$ the value $u_k^2 = 0$, $v_k^2 = 1$ and there 
will be the non-zero spectral function $A_{k \uparrow } \left( E \right) = 
\delta \left( {E - E_{k \uparrow }^ - } \right)$ corresponding to only one 
dispersive state (\ref{eq11}). In Fig.~\ref{fig2}b the spectral functions versus energy for 
different symmetric points in momentum space are shown. Comparison of 
spectral intensities in case of presence and absence of $n_{sf} $ (solid and 
dash-dotted lines) indicates that the second satellite peak appears above 
the main peak in $\left( {\pi / 2,\pi / 2} \right)$ and $\left( {\pi ,0} 
\right)$ points. It is the satellite peak that represents the in-gap state. 
In the $\left( {\pi / 2,\pi / 2} \right)$ point the distance between two 
peaks is proportional to $J$ (see equation (\ref{eq12})) but at $\left( {\pi ,0} 
\right)$ point the distance is proportional to $\left| {J + t'} \right|$ 
(or, generally, $\left| {J_0^B - t_k^A } \right|)$ and will also take place 
even at zero $J$. The last statement emphasizes importance of 
next-nearest-neighbor hoppings $t'$ at low doping. Indeed, if one consider 
$t - J$ model with only nearest-neighbor hoppings then the doped hole will 
not even be able to move without spin fluctuations ($n_{sf} = 0$, see 
equation (\ref{eq11})) and at low $n_{sf} $ the dispersion will be governed by 
$\left( {t_k^A - J_0^B } \right)$ term in equation (\ref{eq10}) but not by nearest 
neighbor hopping $t_k^B $.

Now we will discuss the higher-order corrections to previous results. Two 
main effects are expected:
i) quasiparticle decay and 
ii) renormalization of the real part of the 
self-energy. The main change with introduction of finite quasiparticle 
lifetime will be broadening of spectral peaks. This effect indirectly 
presented in Fig.~\ref{fig2}b where the delta function peaks in spectral function are 
artificially broadened by the Lorentzian. To analyze renormalization of the 
real part of self-energy we will use more rigorous approximation -- 
generalized Hartree-Fock approximation [23]. In this approximation the 
equation of motion for operator $X_f^{0\sigma } $ is renormalized by 
two-sites static correlation functions:
\begin{equation}
\label{eq14}
i\frac{d}{dt}X_f^{0\sigma } = \left[ {\left( {\varepsilon _1 - \mu } \right) 
+ M_{f\sigma } } \right]X_f^{0\sigma } + \sum\limits_g {\tau _{fg,\sigma } 
X_g^{0\sigma } } ,
\end{equation}

\noindent
where $M_{f\sigma } $ and $\tau _{fg,\sigma } $ are the renormalized 
chemical potential (exchange integral) and hopping integrals 
correspondingly:
\begin{eqnarray*}
M_{f\sigma } &=& \sum\limits_g {t_{fg} \left\langle {X_f^{0\bar {\sigma }} 
X_g^{\bar {\sigma }0} } \right\rangle } \\
&-& \sum\limits_g {J_{fg} \left[ 
{\left\langle {X_f^{00} X_g^{\bar {\sigma }\bar {\sigma }} } \right\rangle + 
\left\langle {X_f^{\sigma \bar {\sigma }} X_g^{\bar {\sigma }\sigma } } 
\right\rangle - \left\langle {X_f^{\sigma \sigma } X_g^{\bar {\sigma }\bar 
{\sigma }} } \right\rangle } \right]}, \\
\tau_{fg,\sigma } &=& t_{fg} \left[ {\left\langle {X_f^{00} X_g^{00} } 
\right\rangle + \left\langle {X_f^{00} X_g^{\sigma \sigma } } \right\rangle 
+ \left\langle {X_f^{\sigma \sigma } X_g^{00} } \right\rangle }\right. \\
&+& \left.{ \left\langle 
{X_f^{\sigma \sigma } X_g^{\sigma \sigma } } \right\rangle + \left\langle 
{X_f^{\bar {\sigma }\sigma } X_g^{\sigma \bar {\sigma }} } \right\rangle } 
\right] + J_{fg} \left\langle {X_f^{0\bar {\sigma }} X_g^{\bar {\sigma }0} } 
\right\rangle.
\end{eqnarray*}

It is clear that the equation (\ref{eq14}) has the same linearized form as in 
Hubbard I approximation but with renormalized chemical potential and hopping 
integrals. It means the qualitative results of Hubbard I consideration will 
be the same but quantitatively they may change. Namely, due to 
renormalization of the exchange integral the distance between the in-gap and 
main spectral peaks will be shorter then expected from (\ref{eq12}) and the 
concentration dependence of peak positions will appear due to both 
renormalizations. Meanwhile, the underlying physics of the in-gap state will 
be unchanged and its dispersion will be governed by spin fluctuations.

\section{Conclusion}

It is clear from the spectral function both in the Fig.~\ref{fig1}b and Fig.~\ref{fig2}b that 
there is a pseudogap between the in-gap band and the valence band, for the 
undoped cuprate both bands are occupied and the chemical potential lies 
above the in-gap band. With doping $\mu (x)$ is pinned to the in-gap state 
[16] up to optimal doping. The pseudogap is $k$-dependent. In the $\left( 
{\pi / 2,\pi / 2} \right)$ point of the BZ the value of the gap is $\Delta 
E\left( {\pi / 2,\pi / 2} \right)\sim J\left( {1 - 2n_{sf} } \right)$, while 
in the $\left( {\pi ,0} \right)$ point $\Delta E\left( {\pi ,0} \right)\sim 
\left| {J + t'} \right|\left( {1 - 2n_{sf} } \right)$. We can compare 
results of the $p - d$ model and the $t - J$ model in our solution only in 
the limit $U \to \infty $ ($J \to 0)$, then we get $\Delta E\left( {\pi / 
2,\pi / 2} \right) \to 0$ and $\Delta E\left( {\pi ,0} \right) \to \left| 
{t'} \right|\left( {1 - 2n_{sf} } \right)$ that corresponds to the Fig.~\ref{fig1}. At 
$J \ne 0$ there is the additional contribution to the pseudogap, and we may 
expect the in-gap satellite both in $\left( {\pi / 2,\pi / 2} \right)$ and 
$\left( {\pi ,0} \right)$ points of the BZ.

In conclusion, we have shown that the spin-polaron effect results in the 
formation of the in-gap band above the top of the valence band even in the 
undoped cuprates. The spectral function of the in-gap states has a form of 
small low energy satellite that can be detected by ARPES measurements. The 
most interesting to the ARPES studies are $\left( {\pi ,0} \right)$ and 
$\left( {\pi / 2,\pi / 2} \right)$ points of the BZ. For the hole doped 
cuprates there are two contributions to the in-gap spectral weight: the mean 
field contribution is given by doping concentration x and the 
spin-fluctuation contribution is given by the magnon concentration $2n_{sf} 
$. The latter is temperature dependent resulting in the increasing satellite 
intensity with the temperature growth.

\begin{acknowledgments}
The authors are thankful to Yu.A. Izyumov and V.I. Anisimov for the 
stimulating discussion. This work was supported by RFFI grant 03-02-16124, 
INTAS grant 01-0654, Russian Academy of Science Program ``Quantum 
macrophysics'', Joint Integration Program of Ural and Siberian Branches of 
RAS and Siberian Branch of RAS (Lavrent'yev Contest for Young Scientists).
\end{acknowledgments}

\end{document}